\documentclass{article}
\usepackage{amssymb,amsmath,subfig,color,mathrsfs,graphicx}
\usepackage{pxfonts}
\usepackage[paperwidth=16cm,paperheight=24cm,textwidth=10cm,textheight=10cm,top=2cm,left=2cm,right=2cm,bottom=2cm]{geometry}
\usepackage{hyperref}
\definecolor{refcolor}{rgb}{0.3,0.3,0.7}
\hypersetup{
    pdftitle={Sliding without slipping under Coulomb friction: opening waves and inversion of frictional force},
    pdfauthor={Vladislav A. Yastrebov},
    pdfsubject={Elastodynamic frictional sliding},
    pdfcreator={LaTeX, Kile},
    pdfkeywords={opening wave, stick-slip pulse, inversion of frictional force},
    pdfnewwindow=true,
    colorlinks=true,
    linkcolor=refcolor,          % color of internal links
    citecolor=refcolor,          % color of links to bibliography
    filecolor=refcolor,      % color of file links
    urlcolor=refcolor           % color of external links
}
\usepackage{graphicx}
\usepackage{xspace}
\usepackage{amssymb,amsmath}
\usepackage{marvosym}
\usepackage{color}

\newcommand{\be}{\begin{equation}}
\newcommand{\ee}{\end{equation}}
\newcommand{\bes}{\begin{equation*}}
\newcommand{\ees}{\end{equation*}}

\definecolor{greenvy}{RGB}{0,128,0}
\definecolor{bluevy}{RGB}{0,191,191}
\definecolor{violetvy}{RGB}{191,0,191}
\definecolor{orangevy}{RGB}{255,150,0}

 \title{\Large Sliding without slipping under Coulomb friction:\\opening waves and inversion of frictional force}
 
 \author{\small Vladislav A. Yastrebov\footnote{E-mail: \href{mailto:vladislav.yastrebov@mines-paristech.fr}{vladislav.yastrebov@mines-paristech.fr}}}

 \date{\footnotesize{\it MINES ParisTech, Centre des Mat\'eriaux, CNRS UMR 7633, BP 87, F 91003 Evry, France}}

\begin{document}

\maketitle

\begin{abstract}
An elastic layer slides on a rigid flat governed by Coulomb's friction law. We demonstrate that if the coefficient of friction is high enough, the sliding localizes within stick-slip pulses, which transform into opening waves propagating at intersonic speed in the direction of sliding or, for high Poisson's ratios, at supersonic speed in the opposite direction. This sliding mode, characterized by marginal frictional dissipation, and similar to carpet fold propagation, may result in inversion of the frictional force direction; at longer time intervals the system demonstrates stick-slip behavior. The mechanism is described in detail and a parametric study is presented.
\end{abstract}


\tableofcontents

\newpage
\section{\label{intro}Introduction}

Frictional slip on interfaces, which separate solids with different elastic properties (referred as bi-material interfaces) controls propagation of interfacial mixed mode cracks in composite materials~\cite{coker:2003}, energy dissipation in brake systems~\cite{anderson1990wear,moirot2003ejmas}
and fault slip~\cite{scholz2002b,rice1983pag,kawamura2012rmp}.
Regardless the huge difference in scales between these examples, frictional sliding demonstrates scale invariant properties, which enable to mimic Earth-scale phenomena, like earthquakes, in the laboratory~\cite{xia:2004,ben-david:2010a}. 
However, the scaling laws and the coupling of different mechanisms involved in frictional slip are not yet fully understood and present an active research area in mechanical, physical and geophysical communities~\cite{muser2001prl,carpinteri2005ijss,urbakh2004nature,kawamura2012rmp}.
Difficulties in studying the elastodynamic frictional sliding on bi-material interfaces originates in unknown exact interface laws, complex interaction of the surface and the body waves as well as non-trivial interplay of the involved time and spatial scales. The fact that the slip occurs on the \emph{bi-material} interface implies that the frictional slip locally changes the contact pressure, which is not the case for similar materials. For the Coulomb's friction law, this strong coupling between normal and tangential tractions results in a mathematically ill-posed problem or flutter instability~\cite{renardy:1992,martins1995jva,ranjith:2001} for a wide range of combinations of materials and friction coefficients.
% , which cannot be resolved in the framework of classical continuum models because of the lack of internal lengthscale. 
For example, for material pairs, for which a generalized Rayleigh interface wave does not exist, the frictional slip is well-posed only if the coefficient of friction is below a certain limit: in the case of contact between an elastic half-space and a rigid flat, this limit is a unity~\cite{renardy:1992}. Above this limit the well-posedness depends on the Poisson's ratio $\nu$ and the coefficient of friction $f$~\cite{martins1995jva}. 
% , the stable region vanishes completely for $\nu>1/3$

\begin{figure}[h!]
\includegraphics[width=1\textwidth]{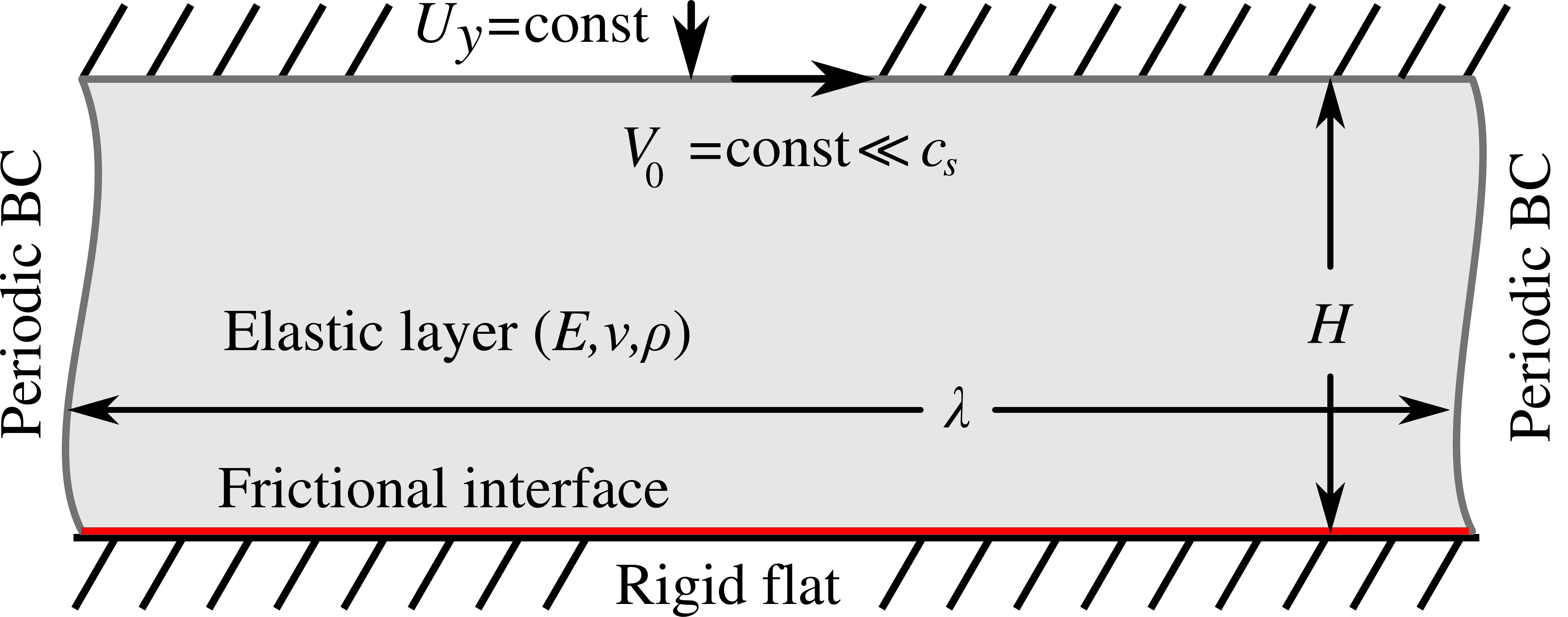}
\caption{\label{fig:1} Problem set-up: infinite elastic layer of thickness $H$ with imposed period $\lambda$ slides on a frictional interface with a rigid flat being sheared at constant rate $V_0$.} 
\end{figure}

In this Letter we demonstrate that the mathematical ill-posedness loses its destructive effect for a finite system, whose frictional dynamics appears to be governed by the structural dynamics.
We report accurate numerical results showing that the frictional sliding in the ``ill-posed'' regime propagates as a series of stick-slip pulses. These pulses, after had run a certain distance, transform into stick-slip-opening waves; both modes of slip propagation imply very low frictional dissipation. The mechanism is described in detail and a parametric study is presented.

\section{Methods}

We consider (Fig.~\ref{fig:1}) an infinitely long elastic layer of height $H$ (Young's modulus $E$, Poisson's ratio $\nu$, mass density $\rho$) pressed against a rigid flat by a constant displacement $U_y<0$ applied on the top and resulting in uniform interface pressure $p$.
The contact interface is governed by Coulomb's friction law with constant coefficient of friction $f$ independent of velocity and slip distance.
From the physical point of view, this implies that at small scale the interface and adjacent materials are assumed to be time independent and not affected by generated frictional heat.
The solid is sheared at a constant velocity $V_0\ll c_s$ applied on the top surface, where $c_s$ is the shear wave celerity.
We use the finite element method with implicit time integration to solve the elastodynamic equation with contact and frictional constraints.

%%%%%%%%%%
%%%%%%%%%%
We solve this problem for an elastic layer of length $\lambda$.
To imitate an infinite elastic layer, periodic boundary conditions were used on lateral sides (displacement degrees of freedom of corresponding nodes are coupled). 
The simulations were carried out in in-house displacement-based finite element software Z-set~\cite{zset,Besson1997}\footnote{For a single case, the simulations were run also in commercial finite element software ABAQUS, which properly reproduced the results obtained with our software.}. To avoid spurious internal wave reflections, we used a regular structured finite element mesh of square-shaped linear elements with reduced integration (one Gauss point per element). To verify the mesh convergence, we used finite elements with the side $h=H/32, H/64, H/128$. Implicit Hilber-Hughes-Taylor~\cite{hilber1977eesd} time integration scheme was used with a moderate dissipation of high frequency modes $\alpha=0.1$. The method establishes the following general time-discrete equations of motion:
$$
  M\ddot u^{t+\Delta t}
  + (1-\alpha) K u^{t+\Delta t} + \alpha K u^t = (1-\alpha) F_{\mbox{\tiny ext}}^{t+\Delta t} + \alpha F_{\mbox{\tiny ext}}^t,
$$
where $M$ is the mass matrix, $K$ is the stiffness matrix, $F_{\mbox{\tiny ext}}$ is the vector of external forces, $u,\dot u, \ddot u$ are vectors of displacement, velocity and acceleration, respectively; the upper index refers to time, and $\Delta t$ is the time step. Plane strain formulation with infinitesimal deformations is used.

We solved elastodynamic equation of motion 
$$
\nabla\!\cdot\!\boldsymbol\sigma = \rho \ddot{\mathbf u},
$$
where $\boldsymbol\sigma$ is the Cauchy stress tensor and $\ddot{\mathbf u}$ is the acceleration vector. The stress-strain relationship is given by Hooke's law
$\boldsymbol\sigma = \frac{E\nu}{(1+\nu)(1-2\nu)} \mbox{tr}(\boldsymbol{\varepsilon})\boldsymbol I + \frac{E}{(1+\nu)} \boldsymbol\varepsilon$, where
$\mbox{tr}(\boldsymbol\varepsilon)$ is the trace of the deformation tensor, and $\boldsymbol I$ is the second rank identity tensor. Since we consider small strain elasticity, the strain tensor is given by
$\boldsymbol \varepsilon= 1/2(\nabla \mathbf u + \mathbf u \nabla)$. The boundary conditions are the following: on the top of the rectangle $y=H$, vertical displacement $u_y=-0.001\,H$ is prescribed, which results in contact pressure $p=U_yE(1-\nu)/[H(1+\nu)(1-2\nu)]$, on lateral sides periodic boundary conditions are imposed for every couple of corresponding nodes $u_x(x=0,y) = u_x(x=\lambda,y)$ and $u_y(x=0,y) = u_y(x=\lambda,y)$, these conditions are kept constant during the simulation. On the contact interface the classic Hertz-Signorini-Moreau contact conditions are used~\cite{wriggers2006b,yastrebov2013book}:
$$
  p\le 0,\quad u_y \ge 0,\quad p u_y = 0
$$
and frictional conditions for Coulomb friction~\cite{wriggers2006b,yastrebov2013book}
$$
  |\sigma_{xy}| - f |p| \le 0,\quad |\dot u_x| \ge 0, \quad (|\sigma_{xy}| - f |p|) \dot u_x = 0,
$$
where $\sigma_{xy}$ is the tangential traction on the contact surface defined as $\sigma_{xy} = \mathbf t \cdot \boldsymbol \sigma \cdot \mathbf n$, where $\mathbf n$ is the unit outward normal and $\mathbf t$ is the unit tangent vector. The frictional contact constraints are taken into account via a direct method, which uses local stiffness matrices to resolve the contact constraints within a nested loop separated from the global convergence loop, in which contact and frictional forces enter as external forces, the method was developed in~\cite{Fra75,Jean1995sam}.

The elastic layer was first loaded by vertical displacement $u_y$ and next the top surface was sheared by $u_x$  until $|T/N|=0.95 f$, where 
$$T = \int\limits_0^\lambda \sigma_{xy} dx,\qquad N = \int\limits_0^\lambda p dx$$ 
are the tangent and the normal forces, respectively; both initial steps are solved quasi-statically (without dynamic effects). Next, a dynamic simulation started (initial velocity field is zero), the top surface is sheared at a constant rate $\dot u_x = V_0$, this rate is chosen to be significantly smaller than the shear wave celerity; for most simulations presented in the Letter we used $V_0 = 10^{-5} c_s$.

In almost all simulations we used Young's modulus $E=1000$ Pa, mass density $\rho = 1000$ kg/m$^3$, $H=25$ m,
Poisson's ratio was varied in the range $\nu \in [0,0.49]$ and the coefficient of friction was varied in the range $f \in [0.5, 2.]$. The time step was chosen as $\Delta t \approx 0.5 \mathrm m/c_s$ To verify the validity of results, various mesh periods were tested $\lambda = 2H, 4H, 6H$, mass density was varied as $\rho=0.1, 10, 1000$ kg/m$^3$, the time step was adjusted proportionally to $\rho^{1/2}$. In all cases the slip dynamics retains all its properties. Different shear rates were also tested $V_0/c_s = 10^{-6}, 5\cdot 10^{-6}, 10^{-5}, 5\cdot 10^{-5}, 10^{-4}$, they can change the slip dynamics at longer time scales, but for moderate shear rates the initial frictional drop and the formation of slip pulses and opening waves were observed for the most of studied parameters. Higher shear rate $V_0/c_s = 10^{-4}$ for some parameters may suppress the inversion of frictional force. Note that the influence of the shear rate was not the focus of the present study and requires a detailed analysis.
%%%%%%%
%%%%%%%

\subsection{\label{sec:remark}Remark on the choice of the time step and the mesh density}

The time step in implicit time integration can be arbitrary large as the method is unconditionally stable, however, too large time steps smooth out rapid wave dynamics leading to slow structural dynamic simulation. But if the time step is chosen sufficiently small, of the order of stable explicit time step (Courant-Friedrichs-Lewy condition), the wave dynamics is finely resolved and for time steps $\Delta t < 1 \mathrm m/c_s$, the choice of the time step does not affect the solution for quasi-stationary regime, where only major slip pulses survive and secondary ones die. 

The mesh size is a more critical parameter, but it affects only the short transitional period from the start of sliding up to the localization of slip pulses. According to~\cite{martins1995jva}, for high enough friction coefficients, the homogeneous slip is unstable and the amplitude of interface displacements grows with an exponent proportional to the wavenumber or, in our case, to the inverse of mesh size. This destabilization may result in localization of stick-slip pulses spaced by a single finite element. However, this instability is only observed within the stabilization cycle before the \emph{finite} size of the system manifests itself. When the system starts to act as a whole, only major slip pulses localize (their number is determined by the ratio of height to period and the slip front celerity), at this stage the mesh size does not change solution characteristics, if a sufficient number of elements is used enabling to capture all relevant fields' variations. We tested meshes with 32, 64 and 128 elements in height, and for the stabilized quasi-periodic solution we could capture quantitatively equivalent results. To summarize, the dependence on mesh and time step exists, it is slightly more subtle than in classic dynamic simulations, but this dependence is eliminated in the stable regime when the finite size of the system manifests itself. Note, that to get rid of any manifestation of the ill-posedness, one may use Prakash-Clifton regularization of frictional law~\cite{prakash:1993,ranjith:2001,cochard2000jgr,kammer2014jmps}. However, we believe that introducing the regularized friction law at the interface will not affect the qualitative results within the stable regime.

\begin{figure}[h!]
\begin{center}
\includegraphics[width=0.75\textwidth]{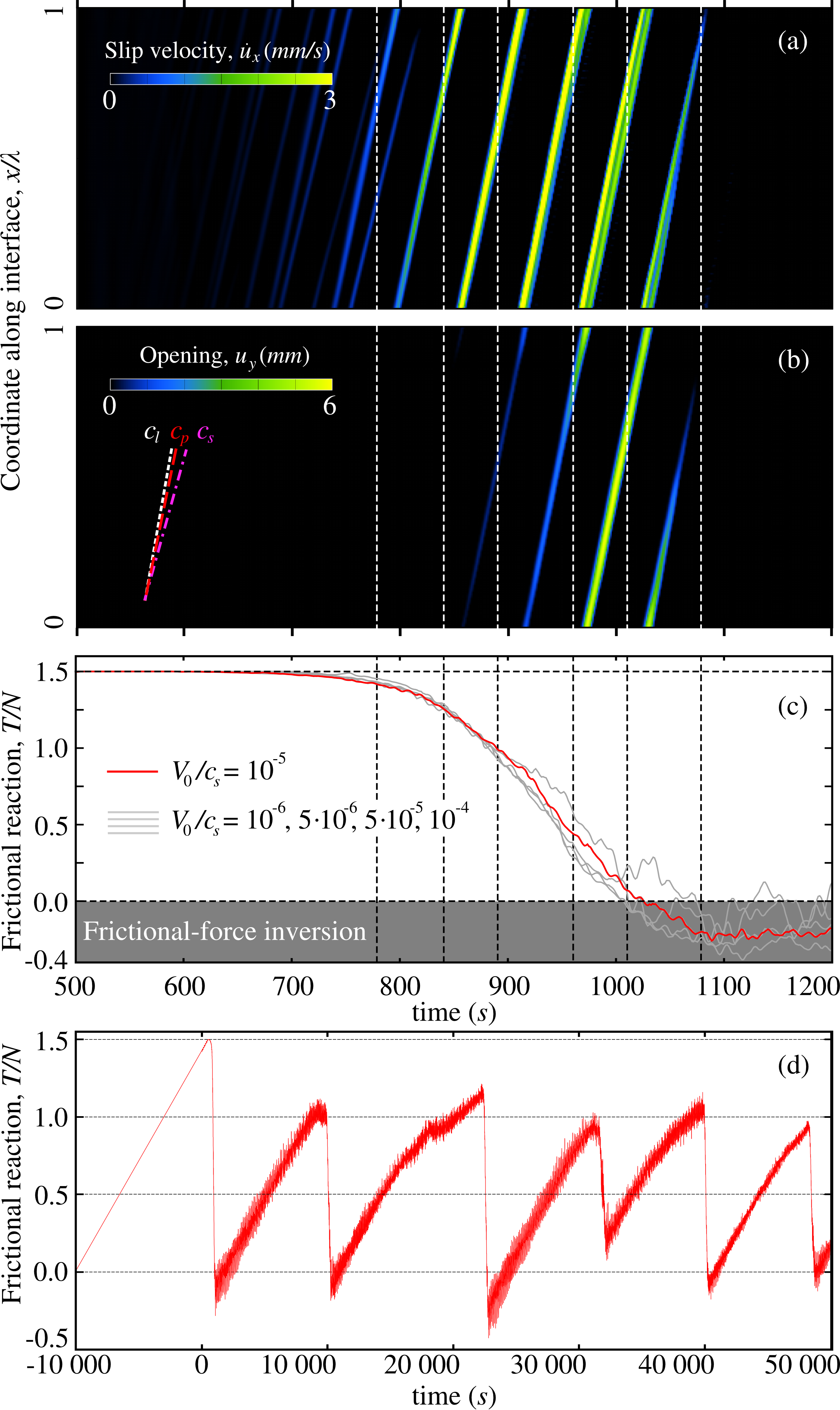}
\end{center}
\caption{\label{fig:2}Spatio-temporal maps showing formation and propagation of (a) slip pulse ($\dot u_x>0$) and (b) opening wave ($u_y>0$), which result (c) in a drop of frictional reaction (ratio of tangential $T$ to normal force $N$ on the interface) down to negative values; longitudinal, shear and slip wave celerities are denoted $c_l,c_s,c_p$, respectively; maps (a,b) and the frictional force (c, red line) are shown for $V_0/c_s=10^{-5}$, frictional forces plotted in gray in (c) are for $V_0/c_s\in[10^{-6};10^{-4}]$. In (d), the ratio $T/N$ for the same simulation is plotted for a longer time interval, $t=0$ corresponds to $T/N\approx0.95 f$.} 
\end{figure}
\section{Mechanism of frictional sliding}

At reaching the critical shear stress $\tau=f|p|$, the frictional interface, initially in stick (pinned) state, starts to slip uniformly. 
But since this uniform slip is unstable for high friction coefficients~\cite{martins1995jva}, after a short transitional period, slip localizes within several slip pulses propagating along the interface (see an example in Fig.~\ref{fig:2} obtained for $\nu=0.2$, $f=1.5$, $E=1000$ Pa, $\rho=1000$ kg/m$^3$, $H=25$ m, $\lambda=50$ m, $V_0/c_s=10^{-5}$).
Material velocity, slip velocity and opening profiles at distinct time moments, which correspond to times marked with white dashed lines in Fig.~\ref{fig:2}, are depicted in Fig.~\ref{fig:3} (an animation of the slip onset and formation of slip and opening pulses is provided in supplemental material~\cite{Supplemental_material}).
Evolution of frictional force for a longer time interval (Fig.~\ref{fig:2},d) is similar to the classical stick-slip behavior. Note, however, that here this global stick-slip dynamics emerges from propagation of rapid slip pulses within time intervals corresponding to frictional drops, in other time intervals the entire interface remains in stick state accumulating elastic energy. Note also that the frictional drops, except the very first one, occur when $T/N\approx 1$ regardless the fact that the local coefficient of friction $f=1.5$.

The observed local propagation mode is referred as train of slip pulses~\cite{adams2000jam,coker:2003,bui2010assm}; solitary pulses in finite systems were also observed~\cite{baumberger2002prl}.
For moderate Poisson's ratios, slip pulses propagate at intersonic celerity $c_p$ ($c_s \!<\! c_p \!<\! c_l$), where $c_l$ is the longitudinal wave celerity. 
Life period of many of these initially formed pulses is short: being filtered by wave dynamics and the resonance of the finite size system, their propagation speed and slip intensity decrease before they arrest in a manner rather similar to what was observed in~\cite{kammer2014jmps}.
However, the major pulses keep going through the interface at the constant speed.
After a certain propagation distance an opening occurs within the slip pulse.
This opening implies that the interface can slide almost without local slipping and thus without any energy dissipation.
One may think of a carpet fold analogy: a carpet can be moved along a surface by simply propagating a fold, which requires much less energy than sliding the entire carpet~\cite{Comninou1978ijss}.
However, here the opening starts at location, where the slip velocity reaches its maximum (Fig.~\ref{fig:3},c-e), thus a marginal frictional dissipation still takes place.
The opening pulses demonstrate a self-sustained behavior and have a certain ``inertia'', so that their propagation often results in bringing the contact surface ahead of the top one. 
This unconventional mechanism results in inversion of the frictional force direction and thus in instantaneous negative apparent friction~(Fig.~\ref{fig:2},c).
It is partly similar to the snap-through instability with accompanied negative dynamic stiffness~\cite{lakes2001n}.
The frictional dissipation is marginal in this slip mode, thus the total energy is approximately conserved. The elastic energy liberated within the first frictional drop via stress waves, is partly restored in the solid being sheared in the opposite direction (Fig.~\ref{fig:3},f). However, in the particular case of the softly applied shear load via a ``pusher''~\cite{ben-david:2010a,radiguet2013prl}, this stored energy will be also liberated.
Such a huge drop in frictional force down to negative values was observed in molecular dynamic simulations~\cite{sorensen1996prb,urbakh2004nature}; also,
within a simple stick-slip model (a rigid block pulling by a spring over a frictional surface under velocity weakening friction with the static and kinetic coefficients of friction $f_s>f_k$) the spring force may change the sign if $f_k < f_s/2$.
Evidently, for systems governed by Coulomb fricton with the coefficient of friction $f$ independent of velocity and slip distance, frictional drop is impossible in the quasi-static case when inertial effects are neglected. For elastodynamic case, it was argued that the apparent kinetic friction (the ratio of the tangential to the normal force) at bi-material interfaces should be smaller than the local one~\cite{adams2000jam} or even vanish completely~\cite{adams1998jam,ranjith:2001}. This is possible due to the drop of contact pressure within the slip pulse  theorized by Weertman~\cite{weertman1980jgr}. However, for the best of our knowledge, the changing of the frictional force direction was not observed at macroscopic scale.

\section{Bibliographical remarks}

The opening waves, also known as Schallamach waves, is a familiar phenomenon in rubber friction~\cite{schallamach1971w,gent1974wear,barquins1985mse}: when a rigid slab slides over a rubber substrate at moderately high velocity, the relative motion between two solids occurs in the form of detachment waves, the energy is dissipated mainly in the bulk of the rubber.
This phenomenon is qualitatively similar to our opening waves, however, the physics of the Schallamach waves, which propagate at smaller speeds than what is observed here, includes visco-elastic material behavior and adhesion on the contact surface, both are not present in our model.
The possibility of detachment waves was also speculated in~\cite{comninou1977jam,Comninou1978ijss} either for similar materials or for frictionless contacts, however, 
some inconsistency in energy balance in these results was pointed out in~\cite{freund1978jam}.
Possibility of detachment was discussed in numerous works~\cite{weertman1980jgr,adams:1995}, but obtaining rigorous theoretical results is associated with great mathematical difficulties.
Nevertheless, the detachment waves on frictional interfaces, different from Schallamach waves, were induced in atomistic simulations~\cite{gerde2001n} and, for the first time, were obtained in finite element simulations for a set-up with rotational symmetry in~\cite{moirot2003ejmas}. 

\begin{figure}[h!]
\includegraphics[width=1\textwidth]{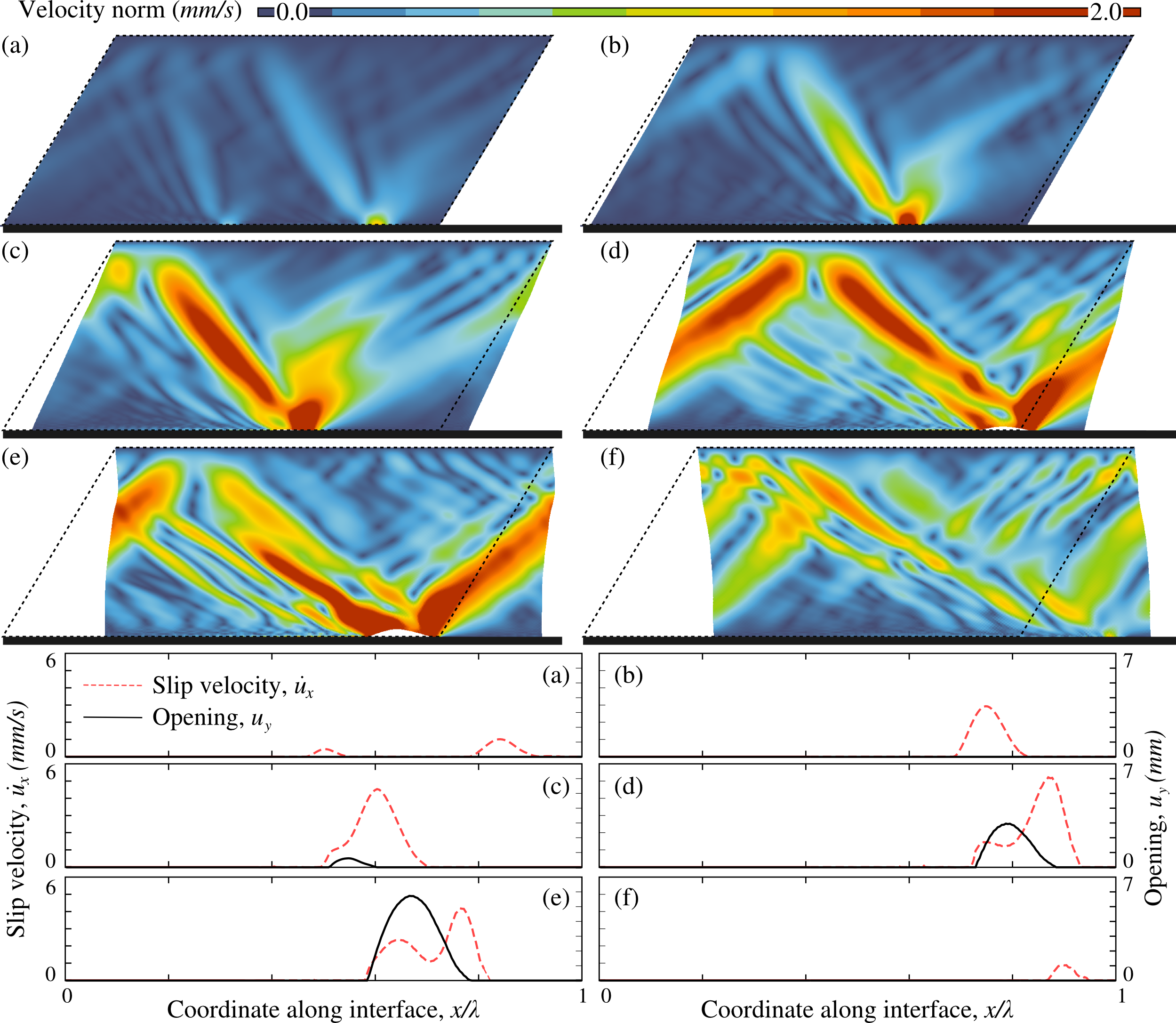}
\caption{\label{fig:3}Instantaneous maps of material velocity in the elastic layer, slip velocity and opening on the interface are presented for $f=1.5$, $\nu=0.2$, $V_0/c_s = 10^{-5}$ at (a) $t\approx787$ s, (b) $t\approx840$ s, (c) $t\approx890$ s, (d) $t\approx960$ s, (e) $t\approx1010$ s, (f) $t\approx1077$ s; deformed state is scaled by a factor $150$; 
dashed line shows the configuration just before the slip starts.}
\end{figure}

\section{Stick-slip and stick-slip-opening pulses}

Here, we consider in detail the formation of an intersonic slip pulse and the mechanism of its self-sustained propagation~\cite{weertman1980jgr,heaton1990pepi,adams:1995}.
The strong coupling between normal and shear tractions on bi-material interface
creates favorable conditions to reach the frictional limit within a local zone and make slip occur and propagate in the direction of sliding of the softer material~\cite{andrews1997jgr,cochard2000jgr,benzion2002jgr}.
Since the study of a solitary pulse in a periodic system is problematic and the number of stable pulses depends on the ratio $\lambda/H$ and on the pulse speed $c_p$, we consider 
a train of stick-slip pulses propagating along the interface at constant speed $c_p$. Let $l$ denote the spacing between pulses and $s$ the width of the pulse. 
Such type of motion was conjectured theoretically~\cite{adams1998jam,adams2000jam,bui2010assm}, observed in experiments~\cite{coker:2003} and in simulations~\cite{cochard2000jgr,coker:2003}.
However, in theoretical findings too restrictive conditions were imposed: the mean slip velocity on the interface was assumed to be equal to the shear velocity $V_0$. This condition is natural if the motion is assumed stationary, it is equivalent to 
the following relation $\dot {\bar u}_x = V_0 l/s$, where $\dot {\bar u}_x$ is the mean slip velocity in the pulse.
Such a stationary solution is observed for low coefficients of friction.
However, such a stationary train of pulses 
cannot result in a huge frictional drop bringing the system in a global pinned state.
Moreover, we observe that the contact surface may be brought ahead of the top one, thus the link between the local and the global speeds should be stated as an inequality
$$\dot {\bar u}_x \ge V_0 l/s\label{eq:1}.$$
Hence, to characterize the train of stick-slip waves, we have four unknowns $c_p,s,l,\dot{\bar u}_x$  and one inequality.

\subsection{Pulse celerity}

As observed in simulations, in most cases the slip front propagates in the sliding direction (direction of the imposed shear) at intersonic velocity $c_p$
(see Fig.~\ref{fig:2} for an example, and Fig.~\ref{fig:4},b and Table~\ref{tab:1} for the summary).
Thanks to the similarity between propagation of the mode-II crack and the slip front~\cite{palmer1973prsl,andrews1976jgr,svetlizky2014n,kammer2014tl}, to find the slip pulse celerity, one may resort to the help of results obtained for intersonic cracks at the interface between an elastic solid and a rigid substrate~\cite{liu1995jmps}. It was shown that the sign of the product between vertical displacement at a certain distance behind the crack tip and the contact pressure ahead of the crack tip is determined by the sign of $(c_p/c_s-\sqrt 2)$. Since the pressure ahead of the slip pulse is always compressive (negative), the vertical displacement is positive if $c_p < \sqrt 2c_s$, thus propagation at this speed is favorable for opening. 
In almost all simulations, in which slip and, in subsequence, opening fronts propagate in the direction of sliding, their celerities were found to be confined in the interval $c_s < c_p < \sqrt 2 c_s$.
The slip front celerity weakly depends on the coefficient of friction, and, in general, the most energetic pulses are obtained for friction coefficients, which result in $c_p \approx \sqrt 2 c_s$.

In Table~\ref{tab:1} slip-pulse celerities $c_p$ are provided for several combinations of parameters, which result in opening waves. Celerity of slip fronts is positive if they propagate in the direction of sliding, and negative if they propagate in the opposite one. The data are obtained for $E=1000$ Pa, $\rho=1000$ kg/m$^3$ and $\lambda=2H$; the average celerity is measured over a time period of $200$ s using a threshold in the slip velocity $\dot u_x$, and was also verified geometrically. The slip celerity remains almost constant when a major slip pulse is localized and does not change when the opening occurs, at least within the precision of simulations and measurements. For $\nu=0$ and $f=2$, in contrast to all other studied combinations of parameters, the slip front propagates in the \emph{sliding direction at supersonic} speed, however, this result does not contradict the theoretical basis~\cite{adams2000jam}. More generally, it is usual in non-linear dynamic systems that for certain combinations of parameters several stable (unstable) solutions coexist, whose realization depends on initial conditions. Probably, here we face a similar situation.

\begin{table}[htb!]
\begin{center}
\resizebox{\textwidth}{!}{
\begin{tabular}{c|ccccccc}
    & \multicolumn{7}{l}{\hspace{2.5cm}Poisson's ratio, $\nu$}\\
  $f$  & 0. & 0.1 & 0.2 & 0.3 & 0.4 & 0.45 & 0.49\\
\hline
  0.50 &	& 	&	&	&	&	& -5.250\\
  0.80 & 	& 	& 	& 	& 	& 	& -5.253\\
  0.90 & 	& 	& 	& 	& 	& -2.506 & \\
  0.95 & 	& 	& 	& 	& 	& -2.523 & \\
  1.00 & 	& 	& 	& 	& 	& -2.523 & \\
  1.05 & 	& 	& 	& 0.851 & 0.825 & -2.511 & -5.248\\
  1.10 & 0.861 	& 0.829 & 	& 0.857 & 	& 	& \\
  1.15 & 0.843 	& 0.824 & 0.880 & 0.858 & 	& 	& \\
  1.20 & 0.861 	& 0.921 & 0.875 &  	& 	& -2.504 & -5.085 \\
  1.30 & 	& 	& 	& 	& 	& 0.708 & \\
  1.40 & 	& 	& 	& 	& 	& 0.708 & \\
  1.50 & 0.982 	& 	& 0.882 & 0.774 & 0.679 & 0.656 & -5.399\\
  1.80 & 	& 0.945 & 0.831 & 0.764 & 0.717 & 	& \\
  2.00 & 1.213	& 	& 	& 0.723 & 0.666 & 0.659 & 0.607\\[2pt]
\hline
  $c_s$ 	& 0.707 & 0.674 & 0.645 & 0.620 & 0.598 & 0.587 & 0.579\\
  $\sqrt 2 c_s$ & 1.000 & 0.953 & 0.913 & 0.877 & 0.845 & 0.830 & 0.819\\
  $c_l$ 	& 1.000 & 1.011 & 1.054 & 1.160 & 1.464 & 1.948 & 4.137\\  [2pt]
\hline
\end{tabular}}
\end{center}
\caption{\label{tab:1}Slip front celerity in (m/s) for various combinations of Poisson's ratios and coefficients of Coulomb's friction. In the three last lines, $c_s,\sqrt 2c_s,c_l$ are provided.}
\end{table}
\subsection{Pulse period}

A relation can be established between $c_p$ and the period length $l$. Since the slip pulse propagates at intersonic celerity, it radiates a shear Mach wave which propagates in the bulk inclined at angle $\theta_s = \arccos(c_s/c_p)$ with respect to the rigid flat. Being reflected from the top surface the Mach wave returns to the frictional interface and might create favorable stress state to initiate or sustain another pulse. Thus $l= 2H\tan(\theta_s) = 2H \sqrt{c_p^2/c_s^2-1}$, and if $c_p\approx\sqrt 2 c_s$ then $l \approx 2H$. This hypothesis was  numerically verified for $f=1.5$, $\nu=0.2$ by varying the ratio $\lambda/H = 2, 4, 6$ resulting in $1,2,3$ major slip-opening pulses. The last unknown, slip length $s$, remains undetermined.

% \subsection{Global behavior}
\subsection{Summary}

\begin{figure}
\includegraphics[width=1\textwidth]{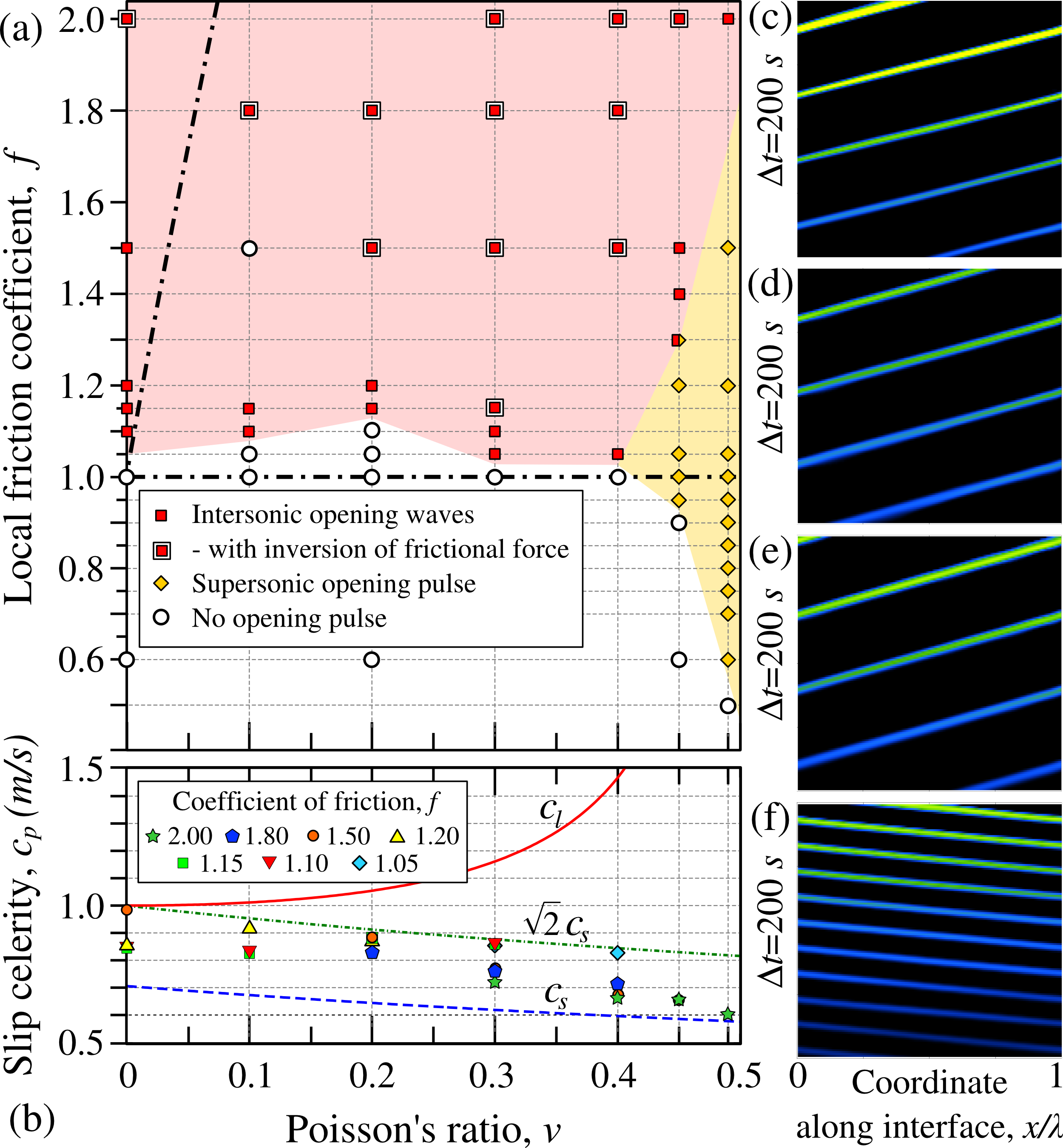}
\caption{\label{fig:4}(a) Intersonic opening waves propagating in the direction of sliding, which are accompanied with high frictional drop, were observed for Poisson's ratios and friction coefficients marked with a square (double square for cases with inversion of frictional force), diamonds mark supersonic opening fronts propagating in the opposite direction, open circles correspond to parameters, for which opening was not observed; parameters corresponding to theoretical flutter instability for slip without opening~\cite{martins1995jva}
is located in the wedge-zone bounded by a dash-dotted line. (b) shows celerities of opening waves for different combinations of Poisson's ratios and friction coefficients; $c_l$, $c_s$ as well as the critical value~\cite{liu1995jmps} $\sqrt 2 c_s$ are also plotted. Examples of slip localization are shown in spatio-temporal maps (black zones correspond to stick state, colored zones correspond to slipping) (c-f) for $(\nu=0,\,f=1.5)$, $(\nu=0.2,\,f=1.1)$, $(\nu=0.3,\,f=1.1)$ and $(\nu=0.45,\,f=1)$, respectively; the latter combination of parameters results in supersonic slip and opening pulses propagating in the direction opposite to sliding.} 
\end{figure}

Opening waves were observed for certain combinations of friction coefficients and Poisson's ratios (Fig.~\ref{fig:4}). Such waves are accompanied with frictional drop (but it does not necessarily leads to inversion of the frictional force). This result holds for all shear velocities in the considered interval $V_0/c_s \in [10^{-6};10^{-4}]$ (see Fig.~\ref{fig:2},c). 
For high Poisson's ratios, within a certain interval of coefficients of friction, slip and opening pulses propagate at supersonic celerity  $c_p>c_l$ in the direction opposite to the sliding; their supersonic celerity is compatible with theoretical results~\cite{adams2000jam}. For higher friction coefficients, slip pulses propagate again in the direction of sliding at intersonic celerity. Seemingly unrealistic, the supersonic slip propagation does not contradict the causality in the system with maximal signal speed equal to the longitudinal wave speed $c_l$. It means simply that the slip advancement is not a result of the local deformation due to the slip, but due to elastic waves traveling in the elastic layer, some experimental and numerical results on supersonic slip can be found in~\cite{coker:2003,coker2005jmps,kammer:2012}.
The time interval of the frictional drop and its value is not affected by the applied shear velocity $V_0$. 
The time step and mesh density affect the slip dynamics only within the transitional phase as discussed in Section~\ref{sec:remark}, but do not alter the formation of major slip pulse and its transformation into opening wave if the time step is small enough to resolve accurately the wave dynamics. 

\section{Conclusion}

Experimental verification of the observed phenomena may be carried out on a set-up with rotational symmetry~\cite{moirot2003ejmas}, which would ensure an initial uniform stress state; non-uniform stress distribution, usual for finite size contacts, results in triggering frictional slip before the local frictional limit $f$ is reached on the global scale. Thus, in finite systems
the apparent or global coefficient of static friction $f_{\mbox{\tiny gl}}^{s}$ should be smaller than the local friction $f$, which is compatible with experimental observations~\cite{ben-david:2010a,ben2011prl}. 
In this light, the behavior of frictional system with a uniform stress distribution resembles to the process of supercooling: the slip may not start before the entire interface reaches the frictional limit, but when it does, the process is rapid and energetic. On contrary, in systems with stress heterogeneities, easily triggered slip front may propagate along the entire interface resulting in global sliding~\cite{ben-david:2010a,kammer2014tl}. 

The sliding mechanism discussed in this Letter adds a new piece to the puzzle of elastodynamic sliding under Coulomb's friction law in finite-size systems, 
but the full picture is still missing especially in the region of high coefficients of friction and Poisson's ratios. 
The prospective study is planned to focus on the dynamics at longer time intervals.

\section{Acknowledgment}
The author is grateful to David S. Kammer and Jean-Pierre Vilotte for valuable discussions.

\newpage

% \bibliographystyle{apalike} 
% \bibliography{references} 
% \input{ref_article_opening_waves.tex} 

%%%%%%%%%%%%%%%%%%%%%%%%%%%%%%%%%%%%%%%%%%%%%%%%%

\end{document}